\documentclass[letterpaper, twocolumn, pra, superscriptaddress]{revtex4-2}
\usepackage{graphicx, amsmath, amssymb, bm, titlesec, mathtools, braket, natbib, dutchcal, xcolor, upgreek, soul}

\begin{document}
	
	\preprint{APS/123-QED}
	
	\title{Measuring the complete set of spatial Schmidt modes of entangled two-photon fields}
	
	\author{Radhika Prasad}
	\email{radhikap@iitk.ac.in}
	\affiliation{%
		Department of Physics, Indian Institute of Technology Kanpur, Kanpur, UP 208016, India}
	\author{Nilakshi Senapati}%
	\affiliation{%
		Department of Physics, Indian Institute of Technology Kanpur, Kanpur, UP 208016, India}
	\author{Abhinandan Bhattacharjee}
	\affiliation{Paderborn University, Warburger Str. 100, 33098 Paderborn, Germany}
	\author{Suman Karan}
	\affiliation{%
		Department of Physics, Indian Institute of Technology Kanpur, Kanpur, UP 208016, India}
	\author{Anand K. Jha}
	\email{akjha@iitk.ac.in}
	\affiliation{%
		Department of Physics, Indian Institute of Technology Kanpur, Kanpur, UP 208016, India}%

	\date{\today}
	
\begin{abstract}

Spontaneous parametric down-conversion (SPDC) is the most widely-used source of high-dimensional entangled two-photon states, and the entanglement in the spatial degree of freedom is considered best suited for harnessing high-dimensional advantages. Although the Schmidt basis provides a natural choice for state characterisation of entangled two-photon states in any degree of freedom, there is currently no technique that can measure the Schmidt basis of an entangled two-photon field. The existing techniques can only reconstruct the Schmidt spectrum when the Schmidt basis is known a priori. In contrast, we present a technique that measures the complete set of spatial Schmidt modes without any prior knowledge. Using this technique, we report measurement of states with over 3000 Schmidt modes---highest reported yet---with up to 98$\%$ fidelity. We expect our work to significantly advance the harnessing of high-dimensional advantages in SPDC-based systems.

\end{abstract}
	
\maketitle
	

{\it Introduction:} High-dimensional entangled states offer increased information capacity \cite{wang2012natphot,bozinovic2013science} and noise resilience \cite{zhu2021avsquantsci,ecker2019prx} that are crucial for quantum information protocols including quantum cryptography \cite{mirhosseini2015njp,bechmann2000prl, cerf2002prl}. 
Spontaneous parametric down-conversion (SPDC) is the most-widely used method for generating high-dimensional pure entangled two-photon states \cite{schneeloch2016jopt, defienne2019pra, karan2020jopt, walborn2010physrep}. Although SPDC photons are rendered entangled in several degrees of freedom, the entanglement in the spatial degrees of freedom \cite{howell2004prl,prasad2024prapp,mair2001nature,karan2023prapp} is considered best suited for harnessing high-dimensional advantages \cite{sephton2023natcomm, kong2023prl}. It is known that for pure two-photon states, the natural basis for measuring and characterizing entanglement as well as maximizing mutual information \cite{miatto2012eurphyjd_yao} is the Schmidt basis \cite{nielsen&chuang2010cup, law2004prl, fedorov2014contphy,ekert1995ajp}. The representation of an entangled state in terms of Schmidt modes is referred to as the Schmidt decomposition, \cite{nielsen&chuang2010cup, straupe2011pra}, and the weightage of the modes are called the Schmidt spectrum. However, due to the lack of efficient techniques for measuring  high-dimensional spatial Schmidt basis, the entanglement in the spatial degree of freedom has been mostly explored in bases that are either measurement-convenient, such as the pixel basis \cite{osullivan2005prl, valencia2020high} or the bases for which the detectors are readily available \cite{hu2018sciadv, brougham2012praatmol, zhang2024advoptphot, kulkarni2017natcomm, karan2025sciadv}.

In the last two decades, there have been several theoretical works on the Schmidt decomposition of SPDC photons. Analysis of high-dimensional spatial entanglement in the Schmidt basis was first presented by Law and Eberly  within certain stringent SPDC phase-matching approximations \cite{law2004prl}. The subsequent works made less stringent approximations \cite{fedorov2009jphysbatmol, miatto2012eurphyjd_ve, miatto2012eurphyjd_yao} and worked out Schmidt decomposition even at propagation distances away from the plane of the crystal producing SPDC \cite{schneeloch2016jopt}. The experimental progress towards measuring the Schmidt decomposition has been very limited.  There are several works for measuring the Schmidt spectrum of entangled two-photon fields when the Schmidt basis is known. For example, experimental techniques have been developed for measuring the Schmidt spectrum in the OAM basis \cite{mair2001nature, kulkarni2017natcomm, karan2023prapp}, and also in the one-dimensional and two-dimensional transverse momentum bases \cite{bhat2024prres,cohen2023avsquantsci, straupe2011pra, straupe2013pra}. Some of these measurement techniques employ single-mode fibers \cite{mair2001nature, straupe2011pra, straupe2013pra} for light collection and suffer from poor mode-dependent coupling efficiency \cite{qassim2014josab, karan2025sciadv}. Many other method employ coincidence detection \cite{mair2001nature, cohen2023avsquantsci, straupe2011pra, straupe2013pra}, which involves large number of sequential measurements as well as long data collection times. As a result, these techniques can measure only a limited number of modes in the chosen bases. Thus, currently, there is no technique that can measure the complete set of Schmidt modes of spatially-entangled two-photon fields produced by SPDC in situation in which there is no prior knowledge of the Schmidt basis.

\begin{figure*}[t!]
	\centering
	\includegraphics[width=2\columnwidth]{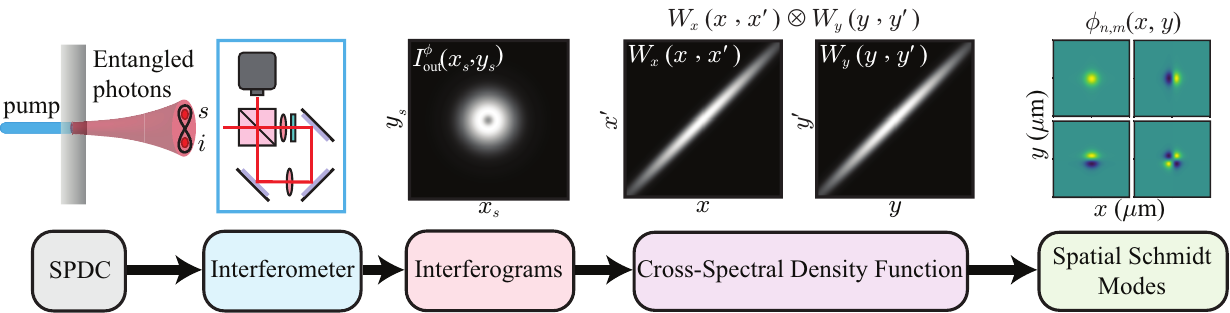}
	\caption{Conceptual illustration of the scheme. One of the SPDC photons (say, the signal photon) goes through an interferometer which measures its cross-spectral density function $W(\bm{\rho}_s,\bm{\rho'}_s)$, which is numerically diagonalized to yield the eigenfunctions $\phi_{mn} (\boldsymbol{\rho}_s)$ and the corresponding eigenvalues $\lambda_{mn}$, which are also the Schmidt modes and Schmidt spectrum of the entangled two-photon state $\psi_{s i}\left(\boldsymbol{\rho}_s, \boldsymbol{\rho}_i\right)$.}\label{fig1}
\end{figure*}

In this letter, we report a technique that measures the complete set of spatial Schmidt modes and Schmidt spectrum of entangled two-photon states without requiring any prior knowledge of the Schmidt basis or involving coincidence counting. The technique  employs a common-path interferometer for measuring the cross-spectral density function of one of the two photons and assumes only the quasi-homogeneity of the individual photon fields.

{\it Theory:} The two-photon wavefunction produced by degenerate SPDC, wherein a pump photon ($p$) with a Gaussian beam profile splits into two photons called the signal ($s$) and idler ($i$) inside a nonlinear crystal of thickness $L$, is given in the transverse momentum basis by \cite{karan2020jopt,schneeloch2016jopt,walborn2010physrep,prasad2024prapp}: 
\begin{equation} \label{eq:mom_state}
\psi_{s i}\left(\boldsymbol{q}_s, \boldsymbol{q}_i\right) \propto \exp \left[\frac{-\left|\boldsymbol{q}_s+\boldsymbol{q}_i\right|^2 w_p^2}{4}\right] \operatorname{sinc}\left[\Delta k_z \frac{L}{2}\right].
\end{equation}
Here $\boldsymbol{q}_s$ and  $\boldsymbol{q}_i$ represent the transverse momenta of the signal and idler photons, $w_p$ is the beam waist of the Gaussian pump beam,  and $\Delta k_z$ is the phase-matching parameter \cite{karan2020jopt,schneeloch2016jopt}. The two-photon wavefunction $\psi_{si}(\bm{\rho}_s,\bm{\rho}_i)$ in the transverse position basis can be obtained by taking the position-space Fourier transform of Eq.~(\ref{eq:mom_state}), which can be written in the Schmidt decomposed form as \cite{nielsen&chuang2010cup,law2004prl,miatto2012eurphyjd_ve,miatto2012eurphyjd_yao,fedorov2009jphysbatmol}:
\begin{equation}
		\psi_{s i}\left(\boldsymbol{\rho}_s, \boldsymbol{\rho}_i\right)=\sum_{m,n} \sqrt{\lambda_{mn}} \phi_{mn}\left(\boldsymbol{\rho}_s\right) \phi_{mn}^{*}\left(\boldsymbol{\rho}_i\right).\label{two-photon-state-position}
\end{equation}
Here $\bm{\rho}_s \equiv (x_s,y_s)$ and $\bm{\rho}_i \equiv (x_i,y_i)$ denote the transverse position coordinates of the signal and idler photons, respectively, $\phi_{mn}(\bm{\rho}_s)$ are the Schmidt modes, and $\sqrt{\lambda_{mn}}$ are the Schmidt coefficients, the square of which is referred to as the Schmidt spectrum ${\lambda_{mn}}$. If the Schmidt decomposition contains only one term, the wavefunction $\psi_{s i}\left(\boldsymbol{\rho}_s, \boldsymbol{\rho}_i\right)$ factorizes, indicating that the two-photon state is separable, that is, not entangled. Therefore, Schmidt-basis analysis provides direct insight into the effective dimensionality and modal structure of an entangled state \cite{law2000prl}. We take the partial trace over the idler modes in Eq.~(\ref{two-photon-state-position}) and obtain the state of the signal photon in terms of the density matrix element: $W(\bm{\rho}_s,\bm{\rho'}_s) = \int \psi_{si}(\bm{\rho}_s,\bm{\rho}_i)
\psi_{si}^*(\bm{\rho'}_s,\bm{\rho}_i)\, d\bm{\rho}_i$, which is also referred to as the cross-spectral density function and can be written as
\begin{align}
W(\bm{\rho}_s,\bm{\rho'}_s)=\sum_{m,n} \lambda_{mn}
		\phi_{mn}(\bm{\rho}_s)\phi^*_{mn}(\bm{\rho'}_s).\label{CMR}
\end{align}
The form of $W(\bm{\rho}_s,\bm{\rho'}_s)$ in Eq.~(\ref{CMR}) is referred to as the coherent mode representation, which represents $W(\bm{\rho}_s,\bm{\rho'}_s)$ as an incoherent sum of modes $\phi_{mn}(\bm{\rho}_s)$ with weights $\lambda_{mn}$. Thus, we find that the eigenfunctions and eigenvalues of the cross-spectral density function of the signal photon are the Schmidt modes and Schmidt spectrum of the entangled two-photon state. Therefore, just by measuring the eigenfunctions and eigenvalues of $W(\bm{\rho}_s,\bm{\rho'}_s)$, one can obtain the complete set of Schmidt modes and Schmidt spectrum of the entangled two-photon field.

\begin{figure*}[t]
	\includegraphics[scale=0.75]{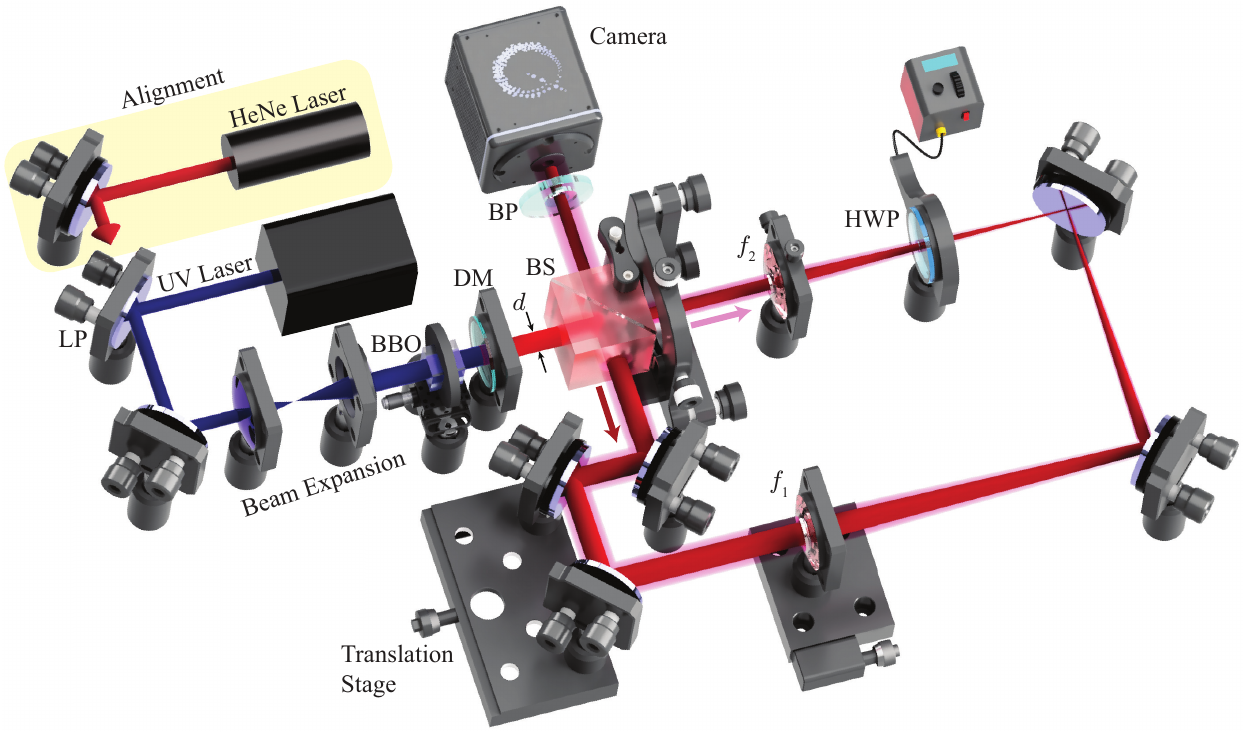}
	\caption{\label{fig:setup}Experimental setup. LP: Longpass filter to transmit HeNe and reflect UV; BBO: $\beta$-Barium Borate crystal; DM: Dichroic mirror to transmit SPDC photons and block UV; BS: Beam Splitter; HWP: Half-Wave Plate; BP: Bandpass filter of bandwidth 10 nm centered at 710 nm. HeNe laser is used only for aligning the interferometer and is turned off when UV laser is used. We represent the interfering alternatives with different shades of dark red for the beam reflected at the BS, and light red for the beam transmitted at the BS.}
\end{figure*}

\begin{figure*}[t]
 	\includegraphics[scale=0.80]{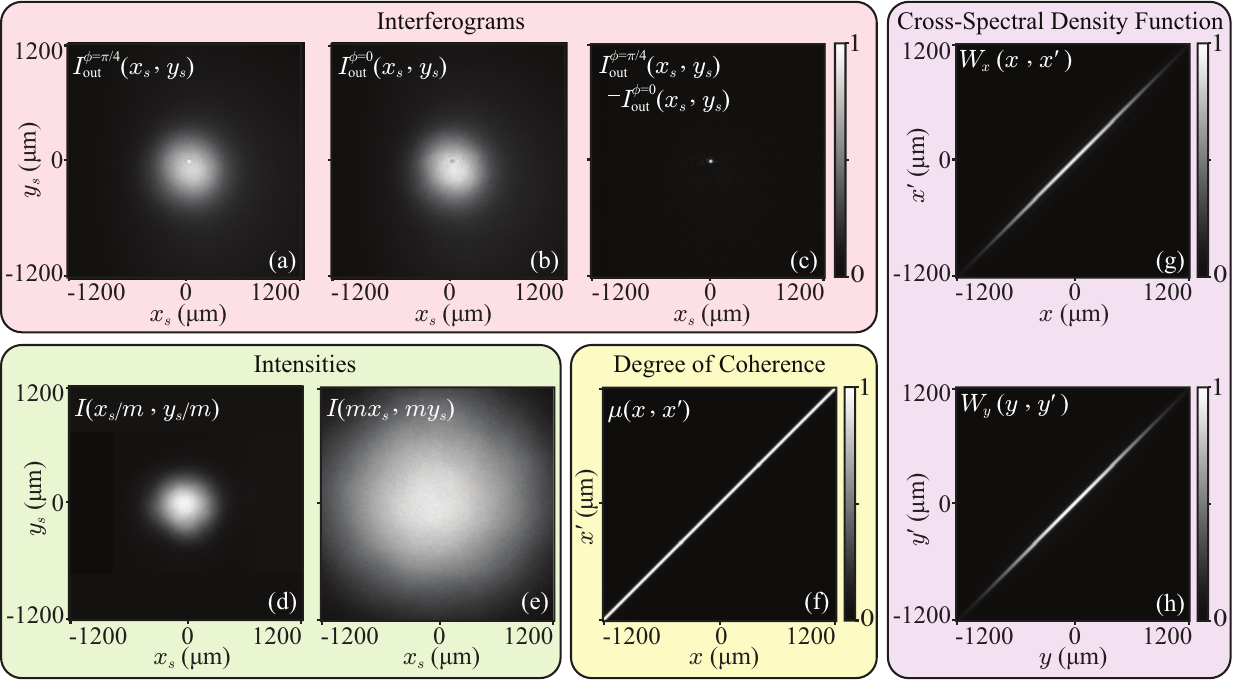}
 	\caption{\label{fig:spec} Experimental observations and reconstructions. (a), (b), (c) are recorded interferograms $I_{\text {out }}^{\phi=\pi / 4}(x_s, y_s)$, $I_{\text {out }}^{\phi=0}(x_s, y_s)$, and their difference. (d) and (e) are the intensities profiles corresponding to the magnified and demagnified alternatives. (f) is the reconstructed degree of coherence function $\mu(x, x')$. (g) and (h) are the reconstructed $W_x(x, x')$ and $W_y(y, y')$.}
\end{figure*}

\label{sec:scheme1}

{\it Measurement scheme:} Figure~\ref{fig1} is the conceptual illustration of our scheme. The experimental technique, involving a Sagnac-type interferometer, is shown in Fig.~\ref{fig:setup} \cite{prasad2025optexp}. The interferometer measures any cross-spectral density function $W_{s}\left(\boldsymbol{\rho}_s, \boldsymbol{\rho}_s^{\prime}\right)$ that is quasi-homogeneous, or in other words, any $W_{s}\left(\boldsymbol{\rho}_s, \boldsymbol{\rho}_s^{\prime}\right)$ that can be expressed as
\begin{equation}
	W(\bm{\rho}_s,\bm{\rho'}_s) = \sqrt{I(\bm{\rho}_s)I(\bm{\rho'}_s)}\mu(\bm{\rho}_s-\bm{\rho'}_s).
	\label{W}
\end{equation}
Here, $I(\bm{\rho}_s)=W(\bm{\rho}_s,\bm{\rho}_s)$ is the transverse position probability of the signal photon and $\mu(\bm{\rho}_s-\bm{\rho'}_s)$ represents the degree of spatial coherence between $\bm{\rho}_s$ and $\bm{\rho'}_s$. The quasi-homogeneity condition requires that (i) the width of $\mu(\bm{\rho}_s-\bm{\rho'}_s)$ must be much smaller than the width of $I(\bm{\rho}_s)$ and (ii) $\mu(\bm{\rho}_s-\bm{\rho'}_s)$ should depend only on the difference coordinate $\bm{\rho}_s-\bm{\rho'}_s$. It can be numerically shown that for type-I, collinear, degenerate SPDC process, $W_{s}\left(\boldsymbol{\rho}_s, \boldsymbol{\rho}_s^{\prime}\right)$ is a quasi-homogeneous function \cite{karan2020jopt}.

The interferometer of Fig.~\ref{fig:setup} consists of $4f$ lens configuration for magnification and a half-wave plate (HWP) for phase control. The field gets magnified by a factor of $m=f_1/f_2$ in alternative 1 (light red, transmitted twice at the beam splitter) and demagnified by the same factor $m$ in alternative 2 (dark red, reflected twice at the beam splitter). The detection probability $I_{\text {out }}^\phi\left(\bm\rho_s\right)$ at the camera plane is given by (for details see Ref.~\cite{prasad2025optexp}): 
\begin{multline}
I_{\text {out }}^\phi\left(\bm\rho_s\right)= I\left(m \bm\rho_s \right) + I\left(\frac{\bm\rho_s}{m} \right)\\-2W\left(m\bm{\rho}_s, \frac{\bm{\rho}_s}{m} \right) \cos (4 \phi)
\end{multline}
where $\bm\rho_s$ is the transverse position at the camera plane and $\phi$ is HWP angle with respect to the horizontal axis. We take the beam splitter to be 50:50 and the cross-spectral density function $W\left(m\bm{\rho}_s, \frac{\bm{\rho}_s}{m} \right)$ to be real. $I\left(m \bm\rho_s\right)$ and  $ I\left(\frac{\bm\rho_s}{m}\right)$  are the detection probabilities of the individual interfering alternatives. We note that $W\left(m\bm{\rho}_s, \frac{\bm{\rho}_s}{m} \right)$ can be measured by recording $I_{\text {out }}^{\phi}(\bm\rho_s)$ at $\phi=0$ and $\phi=\pi/4$, since
\begin{align}
	W\left(m\bm{\rho}_s, \frac{\bm{\rho}_s}{m} \right) = \left[I_{\text {out }}^{\phi=\pi / 4}(\bm\rho_s)-I_{\text {out }}^{\phi=0}(\bm\rho_s)\right].
\end{align}
Next, one needs to measure $I\left(m \bm\rho_s \right)$ and $ I\left(\frac{\bm\rho_s}{m}\right)$. For this, one needs to remove the beam splitter from the interferometer. This way, field from only one interfering alternative reaches the camera and thus one measures $I\left(\frac{\bm\rho_s}{m}\right)$. Since it is a common path interferometer, one cannot directly measure $I\left(m \bm\rho_s \right)$, but it can be assessed by rescaling the measured $I\left(\frac{\bm\rho_s}{m}\right)$ by the factor $m$. The degree of coherence function can therefore be obtained as
\begin{equation}
	\mu\left(m\bm{\rho}_s, \frac{\bm{\rho}_s}{m} \right) = \frac{W\left(m\bm{\rho}_s, \frac{\bm{\rho}_s}{m} \right)}{\sqrt{I(m \bm{\rho}_s)I(\frac{\bm{\rho}_s}{m})}}=\frac{I_{\text {out }}^{\phi=\frac{\pi}{4}}(\bm\rho_s)-I_{\text {out }}^{\phi=0}(\bm\rho_s)}{\sqrt{I(m \bm{\rho}_s)I(\frac{\bm{\rho}_s}{m})}}.
\end{equation}
Now, by utilizing the quasi-homogeneity of the field, we write: $\mu\left(m\bm{\rho}_s, \frac{\bm{\rho}_s}{m} \right) =\mu\left(m\bm{\rho}_s -\frac{\bm{\rho}_s}{m} \right)=\mu\left[\bm\rho_s\left(m-\frac{1}{m}\right)\right]$. Thus, from the experimental measurements, one can obtain $\mu\left(m\bm{\rho}_s-\frac{\bm{\rho}_s}{m} \right)$ for a given value of $m$. We note that the degree of coherence function obtained this way depends only on one variable, $\bm\rho_s\left(m-\frac{1}{m}\right)$. However, the degree of coherence function is a two-point correlation function. So, from the measured values of $\mu\left[\bm\rho_s\left(m-\frac{1}{m}\right)\right]$, one can obtain the two-point degree of coherence function $\mu(\bm\rho, \bm\rho')$ by realizing that the value of $\mu(\bm\rho, \bm\rho')$ for each pair of $(\bm\rho, \bm\rho')$ is related to the single-variable $\mu(\bm\rho-\bm\rho')$ by $\mu(\bm\rho, \bm\rho')=\mu(\bm\rho-\bm\rho')$. Once $\mu(\bm\rho, \bm\rho')$ is numerically reconstructed, one can numerically obtain the cross-spectral density function $W(\bm\rho, \bm\rho')$ using Eq.~(\ref{W}). By diagonalizing $W(\bm\rho, \bm\rho')$, one can finally obtain the coherent modes of Eq.~(\ref{CMR}) and thus the Schmidt modes of Eq.~(\ref{two-photon-state-position}).

{\it Experiment:} Figure~\ref{fig:setup} shows the schematic of the experimental setup. A continuous-wave ultraviolet (UV) laser of wavelength $355$ nm and beam waist $507$ $\mu$m is used to pump a type-I $\beta$-barium borate (BBO) crystal of length $L=1$ mm, which produces collinear SPDC photon pairs. A dichroic mirror placed immediately after the crystal blocks the residual UV pump. A $4f$ imaging system with focal lengths $f_1=400$ mm and $f_2=200$ mm images the crystal plane onto an ORCA-Quest-2 qCMOS camera with a resolution of $4096 \times 2304$ pixels and a pixel size of $4.6\mu\text{m}\times4.6\mu\text{m}$. The imaging system is arranged such that one arm of the interferometer provides a magnification factor of $m=2$, while the other arm produces a magnification factor of $\frac{1}{m}=\frac{1}{2}$ at the camera plane. In the collinear configuration, the signal and idler fields co-propagate through the interferometer, and thus the interferogram recorded by the camera is the sum of the interferograms corresponding to both fields. However, since the both fields have identical spatial distributions, the recorded interferogram is equivalent to that of either the signal or the idler field. A translation stage is used for adjusting the separation between the lenses $f_1$ and $f_2$ to $600$ mm. The half-wave plate (HWP) controls the relative phase $\phi$, and the interferograms are recorded with an acquisition time of 0.5 seconds.

\begin{figure}[t]
	\includegraphics[scale=0.8]{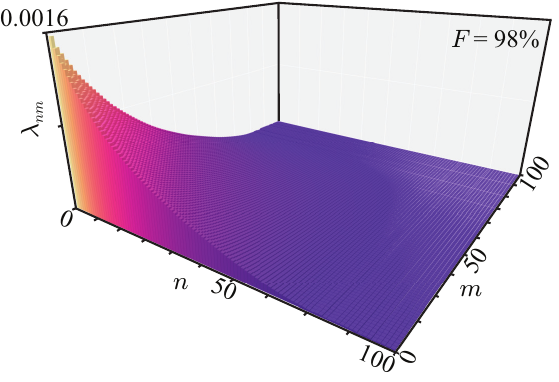}
	\caption{Reconstructed Schmidt spectrum $\lambda_{mn}$. $F$ represents the fidelity of reconstruction.}\label{spectrum}
\end{figure}
\begin{figure*}[t]
	\includegraphics[scale=0.75]{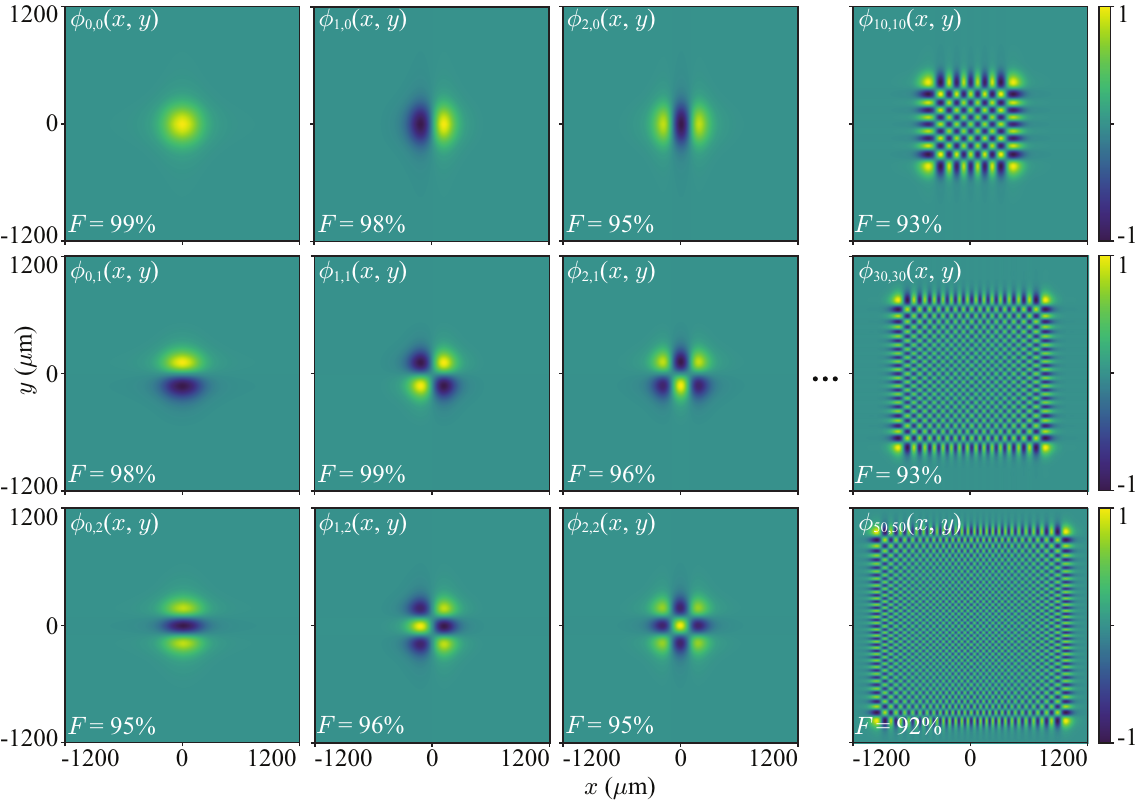}
	\caption{Plots of experimentally reconstructed Schmidt modes $\phi_{mn}(x,y)$. $F$ is the fidelity of reconstruction.}\label{modes}
\end{figure*}

{\it Results:} Figures~\ref{fig:spec} shows our experimental observations. In the figure, the plots have been presented in Cartesian coordinates. Figures~\ref{fig:spec}(a) and \ref{fig:spec}(b) show the recorded interferograms $I_{\text {out }}^{\phi=\pi / 4}(x_s, y_s)$ and $I_{\text {out }}^{\phi=0}(x_s, y_s)$ at $\phi=\pi/4$ and $\phi=0$, respectively. The difference intensity $I_{\text {out }}^{\phi=\pi / 4}(x_s, y_s)- I_{\text {out }}^{\phi=0}(x_s, y_s)$ has been plotted in Fig.~\ref{fig:spec}(c). Figures~\ref{fig:spec}(d) and ~\ref{fig:spec}(e) show the recorded intensity plots $I(\frac{x_s}{m},\frac{y_s}{m})$ and $I(mx_s,my_s)$. Figure~\ref{fig:spec}(f) plots the reconstructed degree of coherence function $\mu(x, x')$ with $m=2$. We note that the plot of $\mu(x, x')$ in Fig.~\ref{fig:spec}(f) depends only on the $x-x'$ coordinate and has no dependence on $x+x'$ and that the width of $\mu(x, x')$ is much narrower than that of $I(x)$, thus verifying that the field of the signal photon is indeed quasi-homogeneous. Finally, from the functions plotted in Figs.~\ref{fig:spec}(d), \ref{fig:spec}(e), and \ref{fig:spec}(f), we reconstruct $W(\bm\rho, \bm\rho')\equiv W_x(x, x')\otimes W_y(y, y')$. Figures~\ref{fig:spec}(g) and ~\ref{fig:spec}(h) show the reconstructed $W_x(x, x')$ and $W_y(y, y')$, respectively. Next, by diagonalizing $W(\bm\rho, \bm\rho')$ in a in a $100\times100$ dimensional space, we obtain the Schmidt spectrum $\lambda_{mn}$ and plot it in Fig.~\ref{spectrum}. Figure \ref{modes} shows the corresponding Schmidt modes $\phi_{mn}(x, x')$ of Eq.~(\ref{two-photon-state-position}). In Fig.~\ref{modes}, we have plotted some representative Schmidt modes up to order 50. We take fidelity $(F)$ as the coefficient of determination \cite{karan2023prapp}. We find that the measurement fidelity for the Schmidt modes in ig.~\ref{spectrum} is about 98\% while for the Schmidt modes in Fig.~\ref{modes} it is up to 99\%. The construction fidelity decreases to about 90\% for modes of the order of 50, which we ascribe to the finite pixel resolution of the detection camera. From the reconstructed spectrum, we determine the effective mode content of the SPDC state, which yields a Schmidt number $K=1/{\sum_{n,m}\lambda_{nm}^2}$ \cite{law2004prl} of approximately $3413$. We note the our measurement fidelity for higher-order Schmidt modes decreases


In conclusion, we have presented a technique for measuring the complete set of spatial Schmidt modes of the high-dimensional entangled two-photon states produced by SPDC without any prior knowledge of the Schmidt basis or requiring coincidence detection. Using this technique, we report measurement of states with over 3000 Schmidt modes and with up to 98$\%$ fidelity. To the best of our knowledge, this represents the largest number of spatial Schmidt modes experimentally measured to date. Our technique can in principle be extended to non-degenerate SPDC in which one of the entangled photon is in the visible range while the other one is in the mid-infra-red range. In such systems, Schmidt modes can be obtained by making measurement only on the visible photons without requiring demanding mid-infra-red detectors. There have been parallel efforts with partial success at developing spatial mode sorters \cite{mirhosseini2013natcomm, berkhout2010prl, sahu2018optexp, mair2001nature, heckenberg1992optlett, karan2020jopt, qassim2014josab}. As a mode sorter allows for the manipulation and control of individual modes in a given basis, our technique, which identifies the Schmidt basis,  combined with an appropriate mode sorter, which would sort the Schmidt modes, could enable not only the measurement and characterization of spatial entangled states but also its manipulation and control for the optimal harnessing of high-dimensional advantages.

{\it Acknowledgments:} We acknowledge financial support from the Science and Engineering Research Board through grants STR/2021/000035 \& CRG/2022/003070, and from the Department of Science \& Technology, Government of India through grant DST/ICPS/QuST/Theme-­1/2019 and through the National Quantum Mission (NQM). R.P. and N.S. thank the Prime Minister’s Research Fellowship, Ministry of Education, Government of India for financial support.

	\bibliography{schmidt2d}

\end{document}